\documentclass[fleqn,twoside]{article}
\usepackage{amsfonts,bezier,epsf}

\newcommand{\bls}[1]{\renewcommand{\baselinestretch}{#1}}
\def\noi{\noindent}

\makeatletter
\renewcommand{\section}{\@startsection{section}{1}{0pt}%
        {-3.5ex plus -1ex minus -.2ex}{2.3ex plus .2ex}%
        {\large\bf\protect\raggedright}}

\renewcommand{\subsection}{\@startsection{subsection}{2}{0pt}%
        {-3ex plus -1ex minus -.2ex}{1.4ex plus .2ex}%
        {\normalsize\bf\protect\raggedright}}

\renewcommand{\thesubsubsection}%
        {\arabic{section}.\arabic{subsection}.\arabic{subsubsection}.}
\newcommand{\para}{\@startsection{paragraph}{4}{0pt}%
        {1.5ex plus -.5ex minus -.2ex}{-1em}{\normalsize\bf}}
\newcommand{\apsection}[2]
	{\subsection*{#1. \ #2}
	\renewcommand{\theequation}{#1.\arabic{equation}}
	\setcounter{equation}{0}
	}

\renewcommand{\@oddhead}{\raisebox{0pt}[\headheight][0pt]{%
   \vbox{\hbox to\textwidth{\rightmark \hfil \rm \thepage \strut}\hrule}}}
\renewcommand{\@evenhead}{\raisebox{0pt}[\headheight][0pt]{%
   \vbox{\hbox to\textwidth{\thepage \hfil \leftmark \strut}\hrule}}}

\newcommand{\heads}[2]{\markboth{\protect\small\it #1}{\protect\small\it #2}}

\makeatother

\newcommand{\Title}[1]{\noi {\Large #1} \\}
\newcommand{\Author}[2]{\noi{\large\bf #1}\\[2ex]\noindent{\it #2}\\}

\newcommand{\Abstract}[1]{\vskip 2mm \begin{center}
        \parbox{16.4cm}{\small\noi #1} \end{center}\medskip}

\newcommand{\foom}[1]{\protect\footnotemark[#1]}

\newcommand{\email}[2]{\footnotetext[#1]{e-mail: #2}}

\newcommand{\Ref}[1]{Ref.\,\cite{#1}}
\newcommand{\sect}[1]{Sec.\,#1}

\def\nq{\hspace*{-1em}}
\def\nqq{\hspace*{-2em}}
\def\nhq{\hspace*{-0.5em}}

\def\cm{\hspace*{1cm}}
\def\inch{\hspace*{1in}}
\def\para{\paragraph}
\newcommand{\Theorem}[2]{\medskip\noi {\bf #1. \ }{\it #2}\medskip}

\newcommand{\Picture}[3]{
	\begin{figure} 	\centering \unitlength=1mm
	\begin{picture}(84,#1)
		\put(0,0){\line(0,1){#1}}            
		\put(0,0){\line(1,0){84}}
		\put(84,0){\line(0,1){#1}}
		\put(0,#1){\line(1,0){84}}
	\put(0,0){#2}                       \end{picture}
        \caption{\protect\small #3}  \smallskip \hrule \end{figure}
	}

\def\eq{Eq.\,}
\def\eqs{Eqs.\,}
\def\beq{\begin{equation}}
\def\eeq{\end{equation}}
\def\bear{\begin{eqnarray}}
\def\al{&\nhq}
\def\lal{&&\nqq {}}               
\def\bearr{\bear \lal}
\def\ear{\end{eqnarray}}

\def\dst{\displaystyle}
\def\tst{\textstyle}

\def\nn{\nonumber\\ {}}

\def\nnn{\nonumber\\ \lal }

\def\yy{\\[5pt] {}}

\def\eql{\al =\al}

\def\eqdef{\stackrel{\rm def}=}
\def\e{{\,\rm e}}
\def\d{\partial}

\def\sign{\mathop{\rm sign}\nolimits}

\def\const{{\rm const}}
\def\Half{{\dst\frac{1}{2}}}
\def\half{{\tst\frac{1}{2}}}

\def\DAL{\mathop{\raisebox{3.5pt}{\large\fbox{}}}\nolimits}

\def\intl{\int\limits}

\def\Jl#1#2{{\it #1\/} {\bf #2},\ }

\def\CQG#1 {\Jl{Class. Qu. Grav.}{#1}}
\def\DAN#1 {\Jl{Dokl. AN SSSR}{#1}}
\def\GC#1 {\Jl{Grav. \& Cosmol.}{#1}}
\def\GRG#1 {\Jl{Gen. Rel. Grav.}{#1}}
\def\JETF#1 {\Jl{Zh. Eksp. Teor. Fiz.}{#1}}
\def\JMP#1 {\Jl{J. Math. Phys.}{#1}}
\def\NPB#1 {\Jl{Nucl. Phys.}{B\ #1}}
\def\PLA#1 {\Jl{Phys. Lett.}{#1A}}
\def\PLB#1 {\Jl{Phys. Lett.}{#1B}}
\def\PRD#1 {\Jl{Phys. Rev.}{D\ #1}}
\def\PRL#1 {\Jl{Phys. Rev. Lett.}{#1}}

\def\GR{general relativity}
\def\sph{spherically symmetric}
\def\ssph{static, spherically symmetric}
\def\fig{Fig.\,}
\def\bh{black hole}
\def\Sch{Schwarzschild}

\def\mn{_{\mu\nu}}
\def\MN{^{\mu\nu}}
\def\mN{_\mu^\nu}

\def\od{{\overline d}}
\def\og{{\overline g}}

\def\M{{\mathbb M}}
\def\R{{\mathbb R}}
\def\S{{\mathbb S}}
\def\oM{{\overline \M}}

\heads{K.A. Bronnikov and G.N. Shikin}
{Spherically symmetric scalar vacuum: no-go theorems, black holes and
solitons}

\bls{0.99}

\begin{document}
\twocolumn[
\thispagestyle{empty}

\rightline{\large\bf gr-qc/0109027}
\bigskip

\Title
{\bf Spherically symmetric scalar vacuum: \yy
	no-go theorems, black holes and solitons}

\Author{K.A. Bronnikov\foom 1 and G.N. Shikin}
{Centre for Gravitation and Fundam. Metrology, VNIIMS,
        3-1 M. Ulyanovoy St., Moscow 117313, Russia;\\
Institute of Gravitation and Cosmology, PFUR,
        6 Miklukho-Maklaya St., Moscow 117198, Russia}

\Abstract
    {We prove some theorems characterizing the global properties of \ssph\
     configurations of a self-gravitating real scalar field $\varphi$ in
     general relativity (GR) in various dimensions, with an arbitrary
     potential $V(\varphi)$, not necessarily positive-definite. The results
     are extended to sigma models, scalar-tensor and curvature-nonlinear
     theories of gravity.  We show that the list of all possible types of
     space-time causal structure in the models under consideration is the
     same as the one for $\varphi = \const$, namely, Minkowski (or AdS),
     Schwarzschild, de Sitter and Schwarzschild --- de Sitter, and all
     horizons are simple.  In particular, these theories do not admit
     regular black holes with any asymptotics. Some special features of
     (2+1)-dimensional gravity are revealed. We give examples of two types
     of asymptotically flat configurations with positive mass in GR, still
     admitted by the above theorems: (i) a black hole with a nontrivial
     scalar field (``scalar hair'') and (ii) a particlelike (solitonic)
     solution with a regular centre; in both cases, the potential
     $V(\varphi)$ must be at least partly negative. We also discuss the
     global effects of conformal mappings that connect different theories
     and illustrate such effects for solutions with a conformal scalar
     field in \GR}

]    
\email 1 {kb@rgs.mccme.ru}

\section {Introduction}

   Vacuum \sph\ solutions to the Einstein equations are either
   Schwarzschild, or, if the cosmological constant is invoked, Schwarzschild
   --- (anti-)de Sitter. All of them, except solutions with zero mass,
   contain curvature singularities at the centre.

   A wider set of space-times is connected with the so-called false vacuum,
   i.e., the system with the action
\beq 							       \label{act}
       S =  \int d^4 x \,\sqrt{-g} [R + (\d\varphi)^2 - 2V(\varphi)]
\eeq
   where $R$ is the scalar curvature, $g=\det (g\mn)$, $\varphi$ is a scalar
   field, $(\d\varphi)^2 = g\MN\d_\mu\varphi\d_\nu\varphi$, and the function
   $V(\varphi)$ is a potential. This action, with many particular forms of
   $V(\varphi)$, is conventionally used to describe the vacuum (sometimes
   interpreted as a variable cosmological term) in inflationary cosmology,
   for the description of growing vacuum bubbles, etc. In space-time regions
   where $\varphi=\const$ (if any), the corresponding potential
   $V(\varphi)$ behaves as a cosmological constant.

   One might expect that the inclusion of a scalar field considerably widens
   the choice of possible qualitative behaviours of \ssph\ configurations.
   Thus, one might seek among them regular \bh\ solutions, trying to
   substantiate the attractive idea of replacing the black hole (BH)
   singularity by a nonsingular vacuum core, which traces back to the 60s
   \cite{glin, bard} but remains in the scope of modern studies. Possible
   manifestations of regular BHs vary from fundamental particles to largest
   astrophysical objects and created universes \cite{dym00,dym01}.

    There are, however, very strong general restrictions that follow directly
    from the Einstein-scalar equations due to (\ref{act}). Thus, if $V\geq
    0$, the only asymptotically flat BH solution is Schwarzschild, as follows
    from the well-known no-hair theorems (see \Ref {bek98} for a recent
    review).  Another result concerns solitonic (particle-like)
    configurations with a regular centre and a flat asymptotic: if $V\geq
    0$, then such a configuration cannot have a positive mass \cite{brsh}.

    An attempt to construct a regular false vacuum BH was made in \Ref
    {hoso1}, with a potential having two slightly different minima,
    $V(\varphi_1) > V(\varphi_2) =0$, the Schwarzschild metric and $\varphi
    \equiv \varphi_2$  outside the horizon, the de Sitter metric and
    $\varphi \equiv \varphi_1$ inside the horizon. It was claimed that a
    reasonable matching of the solutions was possible on the horizon despite
    a finite jump of $\varphi$.  Gal'tsov and Lemos \cite{GaLem} showed that
    the piecewise solution of \Ref{hoso1} cannot be described in terms of
    distributions and requires a singular matter source on the horizon.
    They proved \cite{GaLem} that asymptotically flat regular BH solutions
    are absent in the theory (\ref{act}) with any nonnegative potential
    $V(\varphi)$ (the no-go theorem). Since, by the no-hair theorems,
    $\varphi=0$ outside the horizon, in \Ref{GaLem} it was sufficient to
    show that the \Sch\ exterior cannot be smoothly matched to a regular BH
    interior with $\varphi\ne\const$.

    Less is known if the asymptotic flatness and/or $V\geq 0$ assumptions are
    abandoned. Meanwhile, both assumptions are frequently violated in modern
    studies. Negative potential energy densities, in
    particular, the cosmological constant $V=\Lambda < 0$ giving rise to the
    anti--de Sitter (AdS) solution or AdS asymptotic, do not lead to
    catastrophes (if restricted below), are often treated in various aspects
    and quite readily appear from quantum effects like vacuum polarization.

    We now continue the study of global properties of scalar-vacuum
    configurations begun in \Ref{br01}, which provided some essential
    restrictions on the possible behaviour of solutions of the theory
    (\ref{act}) with arbitrary $V(\varphi)$ in 4 dimensions. This paper
    essentially generalizes and extends the results obtained there.
    \sect 2 presents the field equations in $D$ dimensions, $D\geq 3$.
    \sect 3 contains the proofs of two restriction (no-go) theorems, which
    leave a very narrow spectrum of possible global space-time structures.
    According to Theorem 1, configurations like wormholes and horns are
    absent, while Theorem 3 shows that the variable scalar field adds
    nothing to the list of causal structures known for $\varphi=\const$:
    Minkowski (or AdS), Schwarzschild, de Sitter and Schwarzschild --- de
    Sitter (not to be confused with the de Sitter --- Schwarzschild
    structure having a de Sitter core, discussed in \cite{dym92,dym00}).

    A conclusion much stronger than in \Ref {GaLem}, namely, the
    absence of regular BHs for {\it any\/} $V(\varphi)$ and with {\it
    any\/} asymptotic, then simply follows as a corollary.

    It is shown that in (2+1)-dimensional theory no more than one simple
    horizon is possible, whereas in higher dimensions, including standard
    GR, there can be no more than two simple horizons.

    We also consider the possible existence of geodesically complete
    space-times having curvature singularities at some of their asymptotics
    (the so-called remote singularities). Such singularities at finite
    $r$ are ruled out for any configurations having a regular spatial
    asymptotic (Theorem 2).

    \sect 4 is devoted to 4-dimensional GR: we reproduces the
    no-hair theorem \cite{bek72,AdPear} and the generalized Rosen theorem
    \cite{brsh} for positive-semidefinite potentials.

    \sect 5 discusses extensions of Theorems 1--3 to some more general
    field models. We will consider (i) more general scalar field
    Lagrangians in GR, with an arbitrary dependence on the $\varphi$ field
    and its gradient squared; (ii) multiscalar theories of sigma-model type;
    (iii) scalar-tensor theories of gravity; (iv) curvature-nonlinear
    (high-order) gravity with the Lagrangian of the form $f(R)$ where $f$ is
    an arbitrary function. In items (iii) and (iv), conformal mappings are
    used to reduce the field equations to those due to (\ref{act}) or
    (\ref{act-d}).

    In the Appendix we show that some opportunities allowed by the above
    theorems can be indeed realized by certain choices of $V(\varphi)$.
    Namely, we present particular examples of (i) an asymptotically flat BH
    with positive mass and $\varphi\ne \const$ and (ii) a particle-like
    solution with a regular centre, both being 4-dimensional and having
    positive masses. The potentials for such models are certainly (at least
    partly) negative. In addition, we discuss the well-known solution for a
    conformal scalar field in GR in order to illustrate how conformal
    mappings can affect the space-time global structure.

    To conclude, with all theorems and examples, we now have, even
    without solving the field equations, rather a clear picture of
    what can and what cannot be expected from \ssph\ scalar-vacuum
    configurations in various theories of gravity with various scalar field
    potentials.

    Throughout the paper all statements apply to \ssph\ configurations,
    and all relevant functions are assumed to be sufficiently smooth, unless
    otherwise indicated. The symbol $\DAL$ will mark the end of a proof.

\section{Field equations}

    We start with the $D$-dimensional generalization of the action
    (\ref{act})
\beq 						             \label{act-d}
    S = \int d^D x \,\sqrt{|g|} [R + (\d\varphi)^2 - 2V(\varphi)]
\eeq

    The field equations due to (\ref{act-d}) are
\bear
    \nabla^\alpha \nabla_\alpha \varphi + V_\varphi \eql 0,     \label{SE}
\\
    R\mN -\half \delta\mN\, R + T\mN \eql 0,                    \label{EE}
\ear
    where $V_\varphi \equiv dV/d\varphi$, $R\mN$ is the Ricci tensor and
    $T\mN$ is the energy-momentum tensor of the $\varphi$ field:
\beq
    T\mN = \varphi_{,\mu}\varphi^{,\nu}                         \label{EMT}
                - \half \delta\mN (\d\varphi)^2 + \delta\mN V (\varphi).
i\eeq

    Consider a \ssph\ configuration, with the space-time structure
\beq
    \M^{\od+2} = \R_t \times \R_\rho \times \S^{\od},       \label{stru}
\eeq
    where $\R_t$ is the time axis, $\R_\rho\subset \R$ is the range
    of the radial coordinate $\rho$ and $\S^{\od}$ ($\od=D-2$) is
    a $\od$-dimensional sphere. The metric can be written in the form
\beq                                                          \label{ds-d}
    ds^2 = A(\rho) dt^2 - \frac{d\rho^2}{A(\rho)} - r^2(\rho)
				d\Omega_\od{}^2,
\eeq
    where $d\Omega_\od{}^2$ is the linear element on $\S^{\od}$
    of unit radius, and $\varphi=\varphi(\rho)$. (Without loss of
    generality, we suppose that large $r$ corresponds to large $\rho$.)
    Accordingly, \eq (\ref{SE}) and certain combinations of \eqs (\ref{EE})
    lead to
\bear
       (Ar^\od \varphi')' \eql r^\od V_\varphi;               \label{phi-d}
\\
       	      (A'r^\od)' \eql - (4/\od)r^\od V;               \label{00d}
\\
     	      \od r''/r  \eql -{\varphi'}^2;                 \label{01d}
\\
    A (r^2)'' - r^2 A'' + (\od-2)r'(2Ar' \!\al-\al\! A'r)
\nn
     		\eql 2(\od-1); 				    \label{02d}
\\
   \od(\od-1)(1-A{r'}^2) - \od A'rr' \eql -Ar^2{\varphi'}^2 + r^2 V,
\nnn
							      \label{int-d}
\ear
    where the prime denotes $d/d\rho$. Only three of these five equations
    are independent: the scalar equation (\ref{phi-d}) follows from
    the Einstein equations, while \eq (\ref{int-d}) is a first integral of
    the others. Given a potential $V(\varphi)$, this is a determined set of
    equations for the unknowns $r,\ A,\ \varphi$.

    The choice of the radial coordinate $\rho$ such that
    $g_{tt}g_{\rho\rho} = -1$ is convenient for a number of reasons. First,
    we are going to deal with horizons, which correspond to zeros of the
    function $A(\rho)$. One can notice that such zeros are regular
    points of \eqs (\ref{phi-d})--(\ref{int-d}), therefore one can jointly
    consider regions at both sides of a horizon. Second, in a close
    neighbourhood of a horizon $\rho$ varies (up to a positive constant
    factor) like manifestly well-behaved Kruskal-like coordinates used for
    an analytic continuation of the metric \cite{cold}. Third, with the same
    coordinate, horizons also correspond to regular points in geodesic
    equations \cite{cold}. Last but not least, this
    choice well simplifies the equations, in particular, (\ref{02d}) can be
    integrated, giving, for $\od \geq 2$,
\beq
    B'\equiv \biggl(\frac{A}{r^2}\biggr)' =
	- \frac{2(\od-1)}{r^{\od+2}} \int r^{\od-2} d\rho.      \label{A'd}
\eeq

\section{No-go theorems and global structures}

    Now, our interest will be in the generic global behaviour of the
    solutions and the existence of BHs and globally regular configurations,
    in particular, regular BHs.

    In these issues, a crucial role belongs to {\it Killing horizons,\/}
    regular surfaces where the Killing vector $\d_t$ is null. For the
    metric (\ref{ds-d}), a horizon $\rho=h$ is a sphere of nonzero
    radius $r=r_h$ where $A=0$. The space-time regularity implies the
    finiteness of $T\mN$, so that $V$ and $A {\varphi'}^2$ are finite at
    $\rho=h$. The latter does not forbid $\varphi'\to\infty$ since $A(h)=0$.
    If we, however, additionally (and reasonably) assume that the metric
    functions, including $r(\rho)$, are at least $C^2$-smooth at $\rho=h$,
    then $r''$ is finite, and $|\varphi'| <\infty$ follows from (\ref{01d}).

    The horizon is {\it simple\/} or {\it multiple\/} (or {\it
    higher-order\/}) according to whether the zero of the function $A(\rho)$
    is simple or multiple.  Thus, the Schwarzschild horizon is simple while
    the extreme Reissner-Nordstr\"om one is double.

    As usual, we shall call the space-time regions where $A>0$ and $A<0$
    {\it static\/} (R) and {\it nonstatic\/} (T) regions, respectively.
    A simple or odd-order horizon separates a static region from a nonstatic
    one, whereas an even-order horizon separates two regions of the same
    nature. On the construction of Carter-Penrose diagrams, characterizing
    the causal structure of arbitrary static 2-dimensional space-times [such
    as the ($t,\rho$) section of (\ref{ds-d})] see
    Refs.\,\cite{walker,br79} and more recent and more comprehensive papers
    \cite{katan,strobl}.

\subsection{Regular models without a centre?}

    The first restriction is that such configurations as wormholes,
    horns or flux tubes do not exist under our assumptions.

    For the metric (\ref{ds-d}), a (traversable, Lorentzian) {\it wormhole\/}
    is, by definition, a configuration with two asymptotics at which $r\to
    \infty$, hence with $r(\rho)$ having at least one regular minimum. A
    {\it horn\/} is a region where, as $\rho$ tends to some value $\rho^*$,
    $r(\rho)\ne \const$ and $g_{tt} = A$ have finite limits while the length
    integral $l = \int d\rho/A $ diverges. In other words, a horn is an
    infinitely long $(\od+1)$-dimensional ``tube'' of finite radius, with the
    clock rate remaining finite everywhere. Such ``horned particles'' were,
    in particular, discussed as possible remnants of black hole evaporation
    \cite{banks}.  Lastly, a {\it flux tube\/} is a configuration with $r =
    \const$.

\Theorem{Theorem 1}
    {The field equations due to (\ref{act-d}) do not admit
    (i) solutions where the function $r(\rho)$ has a regular minimum,
    (ii) solutions describing a horn, and
    (iii) flux-tube solutions with $\varphi\ne\const$.
    }

\noi{\bf Proof.} \
    By its geometric meaning (the ``area function'', radius of a
    coordinate sphere), $r(\rho) \geq 0$, therefore \eq (\ref{01d}) gives
    $r''\leq 0$, which rules out regular minima. The same equation leads to
    $\varphi=\const$ as soon as $r = \const$. Thus items (i) and (iii) have
    been proved.

    Suppose now that there is a horn. Then, by the above definition, $A$
    has a finite limit whereas $l\to \infty$ as $\rho\to \rho^*$. This is
    only possible if $\rho^* = \pm \infty$. Under these circumstances,
    the left-hand side of \eq (\ref{A'd}) vanishes at the ``horn end'',
    $\rho\to \rho^* = \pm \infty$, whereas its right-hand side tends to
    infinity.  This contradiction proves item (ii).
$\DAL$

\medskip
    Due to the local nature of the proof, this statement means the absence
    of wormhole throats or horns in solution having any large $r$ behaviour
    --- flat, de Sitter or any other, or having no large $r$ asymptotic at
    all.

    It also follows that the full range of the $\rho$ coordinate covers
    all values of $r$, from the centre ($\rho=\rho_c$, $r(\rho_c)=0$),
    regular or singular, to infinity, unless (which is not excluded) there
    is a singularity at finite $r$ due to a ``pathological'' choice of
    the potential.

    The latter opportunity deserves attention since, being singular at zero
    or finite $r$, the space-time may in principle be still geodesically
    complete. In other words, any geodesic can only reach the singularity
    at an infinite value of its canonical parameter. No freely
    moving particle can then attain such a singularity (to be called a {\it
    remote singularity\/} in finite proper time.
    Examples of remote singularities are known in solutions of 2-dimensional
    gravity \cite{zasl}).

    Let us find out whether such remote singularities can appear in
    systems under consideration. The integral of geodesic equations for
    our metric (\ref{ds-d}) can be written as
\beq                                                         \label{geo}
    \biggl(\frac{d\rho}{d\sigma}\biggr)^2 +kA + L^2\frac{A}{r^2} = E^2,
\eeq
    where $\sigma$ is the canonical parameter, $L$ and $E$ are the conserved
    angular momentum and energy of a moving particle and $k=1,\ 0,\ -1$ for
    timelike, null and spacelike geodesics, respectively. Non-circular motion
    ($r\ne\const$) may be parametrized with the coordinate $\rho$, and the
    canonical parameter can be found from (\ref{geo}) in the form
\beq
    \sigma = \pm\int \frac{d\rho}{\sqrt{E^2-AL^2/r^2-kA}}.  \label{geopar}
\eeq
    This integral can diverge at some $\rho$ for all values of the
    constants of motion $E$ and $L$ simultaneously only if $\rho\to \pm
    \infty$.  Due to $r''\leq 0$, this can happen either at infinity ($r\to
    \infty$), or at finite $\rho=\rho_s$, but then $r < r(\rho_s)$ for all
    values of $\rho$.  We can state the following:

\Theorem{Theorem 2 (on remote singularities).}
    {If a solution to \eqs (\ref{phi-d})--(\ref{int-d}) has a spatial
    asymptotic ($r\to\infty$), it cannot contain a remote singularity at
    $r < \infty$. }

\subsection{Global structures}

    Now, taking into account Theorem 1, the global space-time structure
    corresponding to any particular solution is unambiguously determined
    (up to identification of isometric surfaces, if any) by the
    disposition of static ($A>0$) and nonstatic ($A < 0$) regions.
    The following theorem severely restricts the choice of horizon
    dispositions in the theory under study.

\Picture{78}
{
\unitlength .5mm
\special{em:linewidth 0.4pt}
\linethickness{0.4pt}
\begin{picture}(149.00,136.00)
\put(15.00,65.00){\vector(1,0){132.00}}
\bezier{288}(27.00,126.00)(36.00,95.00)(75.00,86.00)
\bezier{576}(30.00,132.00)(45.00,94.00)(148.00,84.00)
\put(147.00,56.00){\makebox(0,0)[lb]{$\rho$}}
\put(32.00,144.00){\makebox(0,0)[rc]{$B=A/r^2$}}
\put(52.00,11.00){\makebox(0,0)[cc]{1}}
\put(43.00,27.00){\makebox(0,0)[cc]{2}}
\put(37.00,53.00){\makebox(0,0)[cc]{3}}
\put(36.00,101.00){\makebox(0,0)[cc]{4}}
\put(37.00,127.00){\makebox(0,0)[cc]{5}}
\put(20.00,13.00){\vector(0,1){124.00}}
\bezier{260}(75.00,86.00)(106.00,78.00)(130.00,55.00)
\bezier{80}(130.00,55.00)(140.00,47.00)(147.00,46.00)
\bezier{604}(28.00,20.00)(57.00,109.00)(113.00,96.00)
\bezier{148}(113.00,96.00)(137.00,90.00)(149.00,90.00)
\bezier{780}(34.00,11.00)(65.00,118.00)(118.00,53.00)
\bezier{144}(119.00,52.00)(132.00,39.00)(149.00,37.00)
\bezier{512}(47.00,10.00)(69.00,88.00)(102.00,55.00)
\bezier{224}(103.00,54.00)(127.00,30.00)(149.00,29.00)
\end{picture}
}
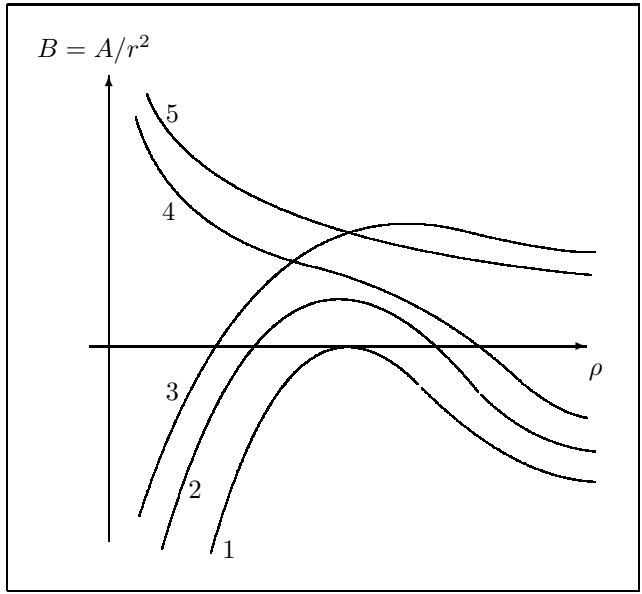
{Generic qualitative behaviours of the function $B(\rho)$ determining the
causal structure of space-time: 1 --- no static region; 2 --- Schwarzschild
-- de Sitter; 3 --- Schwarzschild; 4 --- de Sitter; 5 --- Minkowski/AdS}

\Theorem{Theorem 3 (on the horizons)}
   {Consider solutions of the theory (\ref{act-d}), $D\geq 4$, with the
    metric (\ref{ds-d}) and $\varphi=\varphi(\rho)$. Let there be a static
    region $a < \rho < b \leq \infty$. Then:
\begin{description}\itemsep -2pt
\item [(i)]
    all horizons are simple;
\item [(ii)]
    no horizons exist at $\rho < a$ and at $\rho > b$.
\end{description} }

\noi{\bf Proof.}
    It follows from \eq (\ref{02d}) that if $\rho=h$ is a horizon of order
    higher than 1, i.e., $A(h) = A'(h) =0$, then $A''(h) = -2(\od-1)/r^2
    <0$, i.e., it is a double horizon separating two T regions.  The
    function $I(\rho) = \int r^{\od-2} d\rho$ is monotonically increasing,
    $I'(\rho)>0$, while \eq (\ref{A'd}) implies $I(h)=0$, hence $I<0$ and
    $B'>0$ at $\rho<h$; likewise, $I>0$ and $B'<0$ at $\rho>h$. Therefore
    both $A$ and $B$ are negative for all $\rho\ne h$, i.e., there is no
    static region --- item (i) is proved.

    Consider now the boundary $\rho=a$ of the static region. If $r(a)=0$,
    it is the centre; be it regular or singular, it is then the left
    boundary of the range of $\rho$. If $r(a)\ne 0$, then it is a horizon;
    since $a<\rho<b$ is a static region, $B'(a)>0$, then by (\ref{A'd})
    $I(a)<0$. By monotonicity, $I(\rho{<}a) < 0$, so that $B'(\rho{<}a) >0$.
    This means that $B$ cannot return to zero to the left of $a$, i.e.,
    there is no horizon. In a similar way one can verify that horizons are
    absent to the right of $b$.
    $\DAL$

    According to Theorem 3, there can be no more than two horizons, and
    the list of possible global
    structures is the same as the one well-known for constant $\varphi$:
\begin{description}          \itemsep -2pt
\item[{\rm [TR]:}]
	Schwarzschild/Tangherlini (curve 3 in \fig 1),
\item[{\rm [RT]:}]
	de Sitter (curve 4),
\item[{\rm [R]:}] \
	Minkowski or AdS (curve 5),
\item[{\rm [TRT]:}]
	Schwarzschild -- de Sitter (curve 2), and
\item[{\rm [TT], [T]:}]
	spacetimes without static regions (curve 1 and still below).
\end{description}
    The R and T letters in brackets show the sequence of static and
    nonstatic regions, ordered from center to infinity. The center is
    generically singular. The only possible nonsingular solutions have
    either Minkowski/AdS or de Sitter structures, and, in particular,
    solitonlike asymptotically flat solutions are not excluded.

\Theorem{Corollary}
    {The theory (\ref{act}) does not admit \ssph, regular BHs.}

    Indeed, such a BH, with any large $r$ behavior, must have static
    regions at small and large $r$, separated by at least two simple or
    one double horizon (in the above notation, the structure must be [RTR]
    or [RR] or more complex). This is impossible according to Theorem 3:
    such configurations are simply absent in the above list.

    More generally, if spatial infinity is static, there is at most one
    simple horizon; the same is true if the centre is static.

\subsection*{Special case: (2+1)-dimensional gravity}

    In 3 dimension we have $\od=1$, and integration of (\ref{02d}) leads to
    an expression simpler than (\ref{A'd}):
\beq
	B'\equiv (A/r^2)' = C/r^3, \cm C= \const.    		\label{A'3}
\eeq

    In Theorem 1, items (i) and (iii) hold due to \eq (\ref{01d}), as
    before. Still, the proof of item (ii) does not work: a horn is
    possible if, in (\ref{A'3}), $C=0$. Though, due to $r''<0$, the horn
    radius $r^*$ is the maximum of $r(\rho)$, so that a horned configuration
    has no large $r$ asymptotic.

    By virtue of (\ref{A'3}), $B'$ has a constant sign coinciding with
    $\sign C$, and, instead of Theorem 3, we have a still more severe
    restriction:

\Theorem {Theorem 3a.}
    {A static, circularly symmetric configuration in the theory
    (\ref{act-d}), $D=3$, has either no horizon or one simple horizon.}

    Accordingly, the list of possible global structures is even shorter
    than the previous one: the structures [TT] and [TRT] are now absent, and
    the only structures possessing horizons are \Sch{}-like (possible when
    $C>0$) and de Sitter-like (possible when $C<0$). Purely static and
    purely nonstatic solutions are certainly possible as well; in
    particular, when $C=0$, (\ref{A'3}) gives $A=C_1 r^2$ --- a global R or
    T region depending on the sign of the constant $C_1$.

\section{4-dimensional GR: more restrictions}

    The above theorems did not use any assumptions on the asymptotic
    behaviour of the solutions or the shape and even sign of the potential.
    Let us now mention some more specific but also very significant results
    for positive-semidefinite potentials.

    Consider, for simplicity, $D=4$. The behaviour of the field functions at
    a regular centre and at a flat asymptotic (if, certainly, these are
    present in the solution).

    A {\bf regular centre}, where $r=0$, implies a finite time rate and
    local spatial flatness. This means that at some finite $\rho=\rho_c$
\bear
    	 A{r'}^2 \to 1,   \cm 	  A = A_c + O (r^2),           \label{c1}
\ear
    where $A_c = A(\rho_c)$ and $r'(\rho_c)$ are finite and positive.
    Moreover, the values of $V$, $\varphi$ and $\varphi'$ should be finite
    there. Then from (\ref{01d}) and (\ref{A'd}) one obtains:
\beq
    	  r''(\rho_c) = 0; \cm    \rho_0 = \rho_c.             \label{c2}
\eeq

    At a {\bf flat asymptotic}, the metric should behave as the
    Schwarzschild one with a certain mass $M$, while $\varphi$ should tend
    to a finite value. The corresponding conditions are
\bearr
    \rho\to\infty; \qquad r'\to 1; \qquad
	  		A(\rho) = 1-\frac{2M}{\rho} + O(\rho^{-2});
\nnn
    \varphi' = o (\rho^{-3/2});   \qquad    V = o(\rho^{-3});   \label{a1}
\ear
    The last two requirements follow from the field equations.

    One of the restrictions is the well-known no-hair theorem:

\Theorem{Theorem 4 (no-hair)}
    {Suppose $V \geq 0$. Then the only asymptotically flat BH solution to
    \eqs (\ref{phi-d})--(\ref{int-d}) in the range $(h,\infty)$ (where
    $\rho=h$ is the event horizon) comprises the Schwarzschild metric,
    $\varphi =\const$ and $V\equiv 0$.  }

    This theorem was first proved by Bekenstein \cite{bek72} for the case of
    $V(\varphi)$ without local maxima and was later refined for any
    $V \geq 0$ and for certain more general Lagrangians --- see e.g.
    \Ref{bek98} for proofs and references.

    Let us give, for completeness, a proof similar to that of
    \Ref{AdPear} using our $\rho$ coordinate.

\medskip\noi
{\bf Proof.}
    \eqs (\ref{phi-d}), (\ref{01d}) and (\ref{int-d}) make in possible to
    prove the identity
\beq \nhq
       \frac{d}{d\rho}\biggl[\frac{1}{r'}
       				(2r^2 V - Ar^2{\varphi'}^2)\biggr]
       = 4rV + r{\varphi'}^2                                     \label{id}
			    \biggl(\frac{1}{{r'}^2} + A\biggr).
\eeq

    Suppose that $\rho=h$ is an event horizon and $\rho=\infty$ is flat
    infinity. It follows from (\ref{01d}) that $r(\rho) \leq \rho + \const$.
    Therefore, recalling the horizon properties mentioned in \sect 2, one
    can assert that $h$ is finite while $r(h)=r_h$ and $r'(h)\geq 1$ are
    finite and positive. Integrating (\ref{id}) from $h$ to $\infty$ and
    taking into account (\ref{a1}), one obtains
\beq    \nq
    - \frac{2}{r'(h)}\, r_h^2 V(\varphi(h))
       		= \intl_{h}^{\infty}
\!
        r\biggl[ 4V + {\varphi'}^2                        \label{int-id}
			    \biggl(\frac{1}{{r'}^2} + A\biggr)\biggr]d\rho.
\eeq
    Since here the l.h.s. is nonpositive while both terms in the integrand
    are nonnegative, \eq (\ref{int-id}) only holds if $V$ and
    $\varphi'$ are identically zero for $\rho \in (h,\infty)$. The
    asymptotically flat metric is then necessarily Schwa\-rzschild.
 $\DAL$

    Another restriction can be called the generalized Rosen
    theorem (G. Rosen \cite{rosen} studied similar restrictions for
    flat-space nonlinear field configurations):

\Theorem{Theorem 5 \cite{brsh}}
    {An asymptotically flat solution with positive mass $M$ and a regular
    centre is impossible if $V(\varphi)\geq 0$.          }

\noi {\bf Proof.}
    Let us integrate \eq (\ref{00d}) from the centre ($\rho=\rho_c$)
    to infinity:
\beq
    A'r^2 \biggl|^\infty_{\rho_c} =                          \label{int00}
     			-2 \int_{\rho_c}^{\infty} r^2\, V\, d\rho.
\eeq
    In an asymptotically flat metric, $A(\rho)$ behaves at large $\rho$ as
    $1 - 2M/\rho$, where $M$ is the Schwarzschild mass in geometric units,
    and $r=\rho + O(1)$, therefore the upper limit of $A'r^2$ equals $2M$.
    At a regular centre $r = 0$ and, as is easily verified, $A'=0$, so the
    lower limit is zero. Consequently,
\beq
    M = - \int_{\rho_c}^{\infty} r^2\, V\, d\rho.            \label{V-}
\eeq
    Thus a positive mass $M$ requires an at least partly negative
    potential $V(\varphi)$.
$\DAL$

    $D$-dimensional generalizations of Theorems 3 and 4 are possible but
    will not be considered here.

    The above theorems leave some opportunities of interest, in particular:
\begin{enumerate} \itemsep -1.5pt
\item
    BHs with $\varphi\ne \const$, potentials $V(\varphi)\geq 0$ but
    non-flat large $r$ asymptotics;
\item
    asymptotically flat BHs with $\varphi\ne \const$ but at least
    partly negative potentials $V(\varphi)$;
\item
    asymptotically flat particlelike solutions (solitons) with positive mass
    but at least partly negative potentials $V(\varphi)$.
\end{enumerate}
    That such solutions do exist, one can prove by presenting
    examples. Such examples are known for item 1 due to the paper
    by Chan, Horne and Mann \cite{Mann95}, where, among other results, BHs
    with non-flat asymptotics were found for the Liouville
    ($V=2\Lambda\e^{2b\varphi}$) and double Liouville
    ($V=2\Lambda_1\e^{2b_1\varphi} + 2\Lambda_2\e^{2b_2\varphi}$)
    potentials, where the $\Lambda$'s and $b$'s are positive constants.

    We will give special analytical solutions to \eqs
    (\ref{phi-d})--(\ref{int-d}) for $\od=2$, exemplifying items 2
    (Appendix A)%
\footnote
{Note that two recent examples of exact solutions describing BHs with scalar
hair found by Lechtenfeld and co-authors \cite{lecht} concern scalar fields
with an anomalous sign of kinetic energy, which are of certain interest but
are beyond the scope of this paper.}
    and 3 (Appendix B). Unlike \Ref{Mann95}, where special
    solutions were sought for by making the ansatz $r(\rho)\propto \rho^N$,
    $N=\const$ (in our notation), we will use the following approach.
    Suppose $V(\varphi)$ is one of the unknowns.  Then our set of equations
    is underdeterminate, and we can choose one of the unknowns arbitrarily
    trying to provide the proper behaviour of the solution. Thus, one can
    choose a particular function $r(\rho)$:  assigning it arbitrarily and
    substituting into (\ref{A'd}), by single integration we obtain $A(\rho)$,
    after which $\varphi(\rho)$ and $V(\rho)$ are determined from (\ref{01d})
    and (\ref{00d}), respectively. Thus $V(\varphi)$ is obtained in a
    parametric form; it can be made explicit if $\varphi(\rho)$ resolves
    with respect to $\rho$.

    The purpose of giving these examples is to merely demonstrate the
    existence of such kinds of solutions, therefore the physical meaning of
    the potentials obtained will not be discussed.

\section{Generalizations}

\subsection{More general Lagrangians in GR. Sigma models}

    One can notice that Theorems 1--3 actually rest on two Einstein
    equations, (\ref{01d}) and (\ref{02d}), which in turn follow from
    the properties of the energy-momentum tensor:
    $T^t_t - T^\rho_\rho \geq 0$, which expresses the validity of
    the null energy condition for systems with the metric (\ref{ds-d}).
    This, through the corresponding Einstein equation, leads to $r'' \leq 0$.
    \eq (\ref{02d}), which leads to Theorem 3, follows from the property
\beq
        T^t_t = T^\theta_\theta                              \label{T02}
\eeq
    where $\theta$ is any of the coordinate angles that parametrize the
    sphere $\S^{\od}$.

    Therefore these three theorems hold for all kinds of matter whose
    energy-momentum tensors satisfy these two conditions.

    Consider, for instance, the following action, more general than
    (\ref{act}):
\beq                                                            \label{act'}
	S = \int d^D x \,\sqrt{-g} [R + F(I, \varphi)]
\eeq
    where $I = (\d\varphi)^2$ and $F(I, \varphi)$ is an arbitrary function.
    The scalar field energy-momentum tensor is
\beq
	T\mN = \frac{\d F}{\d I}
         	\varphi_{,\mu}\varphi^{,\nu}                    \label{EMT'}
        	               + \half \delta\mN F (\varphi).
\eeq
    In the \ssph\ case, \eq (\ref{T02}) holds automatically due to
    $\varphi=\varphi(\rho)$, while the null energy condition holds
    as long as $\d F/\d I \geq 0$, which actually means that the kinetic
    energy is nonnegative. Under this condition, all Theorems 1--3 are
    valid for the theory (\ref{act'}). Otherwise Theorem 3 alone holds; it
    correctly describes the $\rho$ dependence of $A$ and consequently
    the possible horizons disposition, but the situation is more complex
    due to possible non-monotonicity of $r(\rho)$.

    Another important and frequently discussed class of theories is
    the class of the so-called sigma models, where a set of $N$ scalar fields
    $\varphi = \{\varphi^a\}$, $a=\overline{1,N}$ are considered as
    coordinates of a target space with a certain metric $G_{ab}=
    G_{ab}(\varphi)$. The scalar vacuum action is then written in the form
\beq \nq
    S_\sigma = \int d^D x \sqrt{|g|}[R +
     		G_{ab}g\MN \d_\mu\varphi^a \d_\nu\varphi^b -2V(\varphi)]
\eeq
    where, in general, $G_{ab}$ and $V$ are arbitrary functions of
    $N$ variables, but in practice they possess symmetries that follow from
    the nature of specific systems.

    It is easily seen that, provided the metric $G_{ab}(\varphi)$
    is positive-definite, Theorems 1--3 for \ssph\ configurations are
    valid as before.

    If $G_{ab}$ is not positive-definite, or if some of $\varphi^a$ are
    allowed to be imaginary, only Theorem 3 holds.

%
%
%
%
%

\subsection{Scalar-tensor and higher-order gravity}

    Other extensions of the present results concern theories connected with
    (\ref{act-d}) and (\ref{act'}) via $\varphi$-dependent conformal
    transformations, such as theories with nonminimally coupled scalar
    fields (e,g., scalar-tensor theories, STT) and nonlinear gravity (e.g.,
    with the Lagrangian function $f(R)$).

    Above all, it should be noted that if a space-time $\M[g]$ with the
    metric (\ref{ds-d}) is conformally mapped into another space-time
    $\oM[\og]$, equipped with the same coordinates, according to the law
\beq
    g\mn = F(\rho) \og\mn,                                   \label{g-conf}
\eeq
    then it is easily verified that
    a horizon $\rho=h$ in $\M$ passes into a horizon of the same
    order in $\oM$, (ii) a centre ($r=0$), an asymptotic ($r\to \infty$)
    and a remote singularity in $\M$ passes into a center, an asymptotic
    and a remote singularity, respectively, in $\oM$
    if the conformal factor $F(\rho)$ is regular (i.e., finite, at least
    C${}^2$-smooth and positive) at the corresponding values of $\rho$.
    A regular centre passes into a regular centre and a flat asymptotic
    to a flat asymptotic under evident additional requirements, but we will
    not concentrate on them here.

    The general (Bergmann-Wagoner-Nordtvedt) STT action in $D$ dimensions
    can be written as follows:
\bearr
     S_{\rm STT} = \int d^D x \sqrt{|g|}                 \label{act-J}
		   [f(\phi) R
\nnn \inch
		   + h(\phi) (\d\phi)^2 -2U(\phi) + L_m],
\ear
    where $f$, $h$ and $U$ are functions of the scalar field $\phi$ and
    $L_m$ is the matter Lagrangian. The metric $g\mn$ here corresponds to
    the so-called Jordan conformal frame. The standard transition to the
    Einstein frame \cite{wagon},
\bear
    g\mn \eql F(\varphi) \og\mn,\cm  F=f^{-2/(D-2)},    \label{g-wag}
\\
    \frac{d\varphi}{d\phi} \eql \frac{\sqrt{|l(\phi}|}{f(\phi},
\qquad                                                     \label{phi-wag}
    l(\phi) \eqdef fh + \frac{D-1}{D-2}\biggl(\frac{df}{d\phi}\biggr)^2,
\ear
    removes the nonminimal scalar-tensor coupling express\-ed in a
    $\phi$-dependent coefficient before $R$. Putting $L_m=0$ (vacuum), one
    can write the action (\ref{act-J}) in terms of the new metric $\og\mn$
    and the new scalar field $\varphi$ as follows (up to a boundary term):
\beq
     S_{\rm E} = \int d^D x \sqrt{|\og|}                     \label{act-E}
	[R_{\rm E} + \eta_l(\d\varphi)^2 -2V(\varphi)],
\eeq
    where $R_{\rm E}$ and $(\d\varphi)^2$ are calculated using $\og\mn$,
\beq
       V(\varphi) = \eta_f F^2 (\varphi)\, U(\phi),           \label{V-E}
\eeq
    and $\eta_{l,f}$ are sign factors:
\beq
     \eta_l = \sign l(\phi), \cm \eta_f = \sign f(\phi).      \label{etas}
\eeq

    Note that $\eta_l = -1$ corresponds to the so-called anomalous STT,
    with a wrong sign of scalar field kinetic energy, while $\eta_f=-1$
    means that the effective gravitational constant in the Jordan frame is
    negative. So the normal choice of signs is $\eta_{l,f}=1$.

    The action (\ref{act-E}) obviously coincides with (\ref{act-d}) up to the
    coefficient $\eta_l$. Therefore \eq (\ref{T02}) holds, and we can assert
    that, for \ssph\ configurations, Theorem 3 is valid for the
    Einstein-frame metric $\og\mn$.

    Theorems 1 and 2 hold for $\og\mn$ only in the ``normal'' case
    $\eta_l=1$; let us adopt this restriction.

    The validity of the theorems for the Jordan-frame metric $g\mn$ depends
    on the nature of the conformal mapping (\ref{g-wag}) between the
    space-times $\M [g]$ (Jordan) and $\oM [\og]$ (Einstein). There are
    four variants:

\def\map{\ \longleftrightarrow\ }
\begin{description}\itemsep -2pt
\item[I.]
	$\M \ \map\ \oM$,
\item[II.]
	$\M \ \map\ (\oM_1 \subset \oM$),
\item[III.]
	$(\M_1 \subset \M) \ \map\ \oM$,
\item[IV.]
	$(\M_1 \subset \M) \ \map\ (\oM_1 \subset \oM$),
\end{description}
    where $\map$ denotes a diffeomorphism preserving the metric signature.
    The last three variants are possible if the conformal factor $F$
    vanishes or blows up at some values of $\rho$, which then mark the
    boundary of $\M_1$ or $\oM_1$.

    Theorem 3 on horizon dispositions is obviously valid in $\oM$ in
    cases I and II. In case III or IV, the whole space-time $\oM$ or its
    part is put into correspondence to only a part $\M_1$ of $\M$, and,
    generally speaking, anything, including additional horizons, can appear
    in the remaining part $\M_2=\M \setminus \M_1$ of the Jordan-frame
    space-time.

    Theorem 1 cannot be directly transferred to $\M$ in any case except the
    trivial one, $F=\const$. It is only possible to assert, without
    specifying $F(\varphi)$, that wormholes as global entities are impossible
    in $\M$ in cases I and II if the conformal factor $F$ is finite in the
    whole range of $\rho$, including the boundary values. Indeed, if we
    suppose that there is such a wormhole, it will immediately follow that
    there are two large $r$ asymptotics and a minimum of $r(\rho)$
    between them even in $\oM$, in contrast to Theorem 1 which is valid
    there.

    Theorem 2 also evidently holds in $\M$ in cases I and II if the
    conformal factor $F$ is regular in the whole range of $\rho$, including
    the boundary values.

    Another class of theories conformally equivalent to (\ref{act-d}) is
    the so-called higher-order gravity with the vacuum action
\beq
     S_{\rm HOG} = \int d^D x \sqrt{|g|}f(R)               \label{act-hog}
\eeq
    where $f$ is a function of the scalar curvature $R$ calculated for the
    metric $g\mn$ of a space-time $\M$. The conformal mapping $\M[g] \mapsto
    \oM[\og]$ with
\beq
    g\mn = F(R) \og\mn,\cm  F= f_R^{-2/(D-2)}, \label{g-hog}
\eeq
    (where, in accord with the weak field limit $f\sim R$ ar small $R$,
    we assume $f(R) >0$ and $f_R \eqdef df/dR >0$),
    transforms the action (\ref{act-hog}) into (\ref{act-d}) with
\bear                                                      \label{phi-hog}
    \varphi\eql \Half\sqrt{\frac{D-1}{D-2}}\log f_R,
\\
    V(\varphi) \eql f_R^{-D/(D-2)} (Rf_R - f). 	              \label{V-hog}
\ear
    The field equations due to (\ref{act-hog}) after this substitution
    turn into the field equations due to (\ref{act-d}).

    All the above observations on the validity of Theorems 1--3
    in STT equally apply to curvature-nonlinear gravity.

    It should be noted that most of papers which use such conformal
    mappings assume, explicitly or implicitly, a one-to-one correspondence
    between $\M$ and $\oM$ (variant I). Meanwhile, the most general
    case is given by variant IV, where the mapping only connects
    subsets of the respective manifolds. This situation is similar to a
    transformation between coordinate frames in a single manifold.

    An exactly soluble example with a conformally coupled scalar field in
    GR, when the mapping follows variant III and there appear horizons or
    wormhole throats outside $M_1$, is presented in Appendix C.

    To conclude, with all these theorems and examples, we now have,
    without solving the field equations, a more or less clear picture of
    what can and what cannot be expected from \ssph\ scalar-vacuum
    configurations in various theories of gravity with various scalar field
    potentials.

\section* {Appendix}
\apsection {A} {An asymptotically flat \bh}

\def\ox{{\bar x}}

    Let us introduce an arbitrary constant length scale $a$ and
    the dimensionless quantities
\beq
    x = \rho/a, \qquad R(x) = r/a; \qquad x_0 = \rho_0/a.     \label{dimls}
\eeq
    The quantities $\varphi$ and $A$ are already dimensionless and are
    expressed in terms of the known (prescribed) function $R(x)$ as follows:
\bear
    \psi(x) \eqdef \varphi/\sqrt{2} = \pm \int \sqrt{-\frac{R_{xx}}{R}}dx,
							       \label{phi-x}
 \\                                                            \label{A-x}
     A(x) = 2R^2 \int_x^\infty \frac{\ox-\ox_0}{R^4(\ox)}\,d\ox.
\ear
    where the subscript $x$ denotes $d/dx$. \eq (\ref{A-x}) is written in
    such a form that $A(\infty)=1$ if $R(x)$ behaves at large $x$ as
    $x + \const + o(1)$, thus providing asymptotic flatness.

    Choosing $R(x)$ in the form
\beq
	R(x) = \sqrt{x^2 - 1},                                 \label{R-bh}
\eeq
    we obtain (restricting to the $+$ sign in (\ref{phi-x})):
\bear
    \psi(x)\eql{\rm Arcoth}\,x \equiv\Half\log\frac{x+1}{x-1},\label{psi-bh}
\\
    A(x)\eql 1 - xx_0 + x_0\frac{x^2-1}{2}\log\frac{x+1}{x-1}. \label{A-bh}
\ear
    Substituting these expressions into (\ref{00d}) and finally inserting
    $x=\coth \psi$, we arrive at an explicit form of the dimensionless
    potential $U=a^2 V$:
\bear                                                          \label{U-bh}
    U \eql\frac{x_0}{2(x^2-1)}\biggl[6x + (1-3x^2) \log\frac{x+1}{x-1}\biggr]
\nn     \eql
     x_0\biggl[\frac{3}{2}\sinh(2\psi)-\psi\cosh(2\psi) - 2\psi\biggr].
\ear

    This is a BH solution in case $x_0 > 0$, and there is a naked singularity
    in case $x_0 <0$. It is easy to verify that the \Sch\ mass $M$
    (according to the asymptotic form $A = 1 -2M/\rho +...$ at large $\rho$)
    is given by $3M = \rho_0 = ax_0$. Therefore the BH branch of this
    solution corresponds to a positive mass, just as in the \Sch\ solution.

    The field $\varphi$ ranges from zero at infinity to infinity at the
    singular centre, $x=1$. There is one simple, \Sch-like horizon ($A=0$),
    in full agreement with the general treatment of \sect 2.

\Picture{82}
{
\unitlength=0.50mm
\special{em:linewidth 0.4pt}
\linethickness{0.4pt}
\begin{picture}(147.00,152.00)
\put(26.00,60.00){\vector(0,1){92.00}}
\put(26.00,152.00){\vector(0,0){0.00}}
\put(22.00,100.00){\vector(1,0){125.00}}
\put(28.00,98.00){\makebox(0,0)[lt]{1}}
\put(46.00,98.00){\makebox(0,0)[ct]{2}}
\put(66.00,98.00){\makebox(0,0)[ct]{3}}
\put(86.00,98.00){\makebox(0,0)[ct]{4}}
\put(106.00,98.00){\makebox(0,0)[ct]{5}}
\put(23.00,115.00){\makebox(0,0)[rc]{0.5}}
\put(23.00,130.00){\makebox(0,0)[rc]{1}}
\put(23.00,145.00){\makebox(0,0)[rc]{1.5}}
\put(23.00,85.00){\makebox(0,0)[rc]{-0.5}}
\put(23.00,70.00){\makebox(0,0)[rc]{-1}}
\bezier{448}(29.00,145.00)(40.00,110.00)(115.00,105.00)
\bezier{456}(29.00,85.00)(40.00,120.00)(117.00,125.00)
\bezier{216}(34.00,61.00)(38.00,99.00)(54.00,100.00)
\put(35.00,139.00){\makebox(0,0)[lb]{$\psi(x)$}}
\put(99.00,128.00){\makebox(0,0)[cb]{$A(x)$}}
\put(40.00,74.00){\makebox(0,0)[lb]{$U(x)$}}
\put(46.00,10.00){\vector(0,1){49.00}}
\put(46.00,50.00){\vector(1,0){82.00}}
\put(48.00,52.00){\makebox(0,0)[lb]{0}}
\put(71.00,52.00){\makebox(0,0)[cb]{0.5}}
\put(96.00,52.00){\makebox(0,0)[cb]{1}}
\put(121.00,52.00){\makebox(0,0)[cb]{1.5}}
\put(44.00,35.00){\makebox(0,0)[rc]{-1}}
\put(44.00,20.00){\makebox(0,0)[rc]{-2}}
\bezier{372}(52.00,50.00)(103.00,52.00)(113.00,11.00)
\put(71.00,35.00){\makebox(0,0)[cc]{$U(\psi)$}}
\end{picture}
}
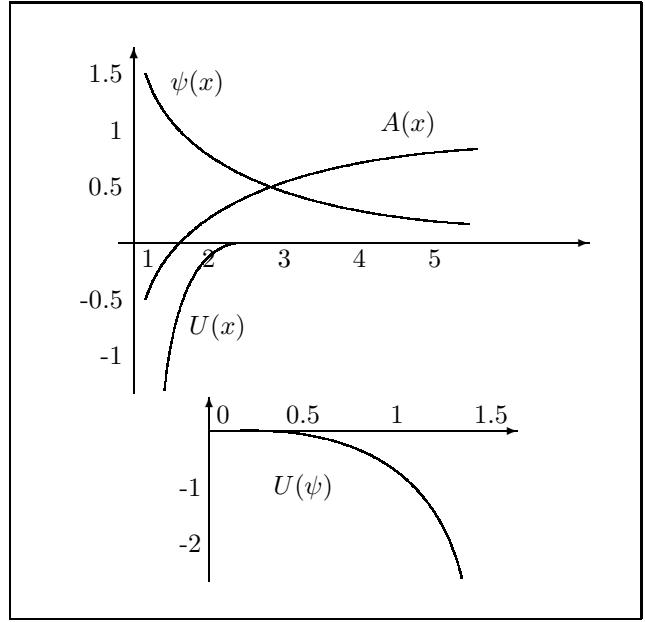
{The functions $\psi(x)$, $A(x)$, $U(x)$ and  $U(\psi)$
 for the example of a \bh\ solution with $x_0=2$.}

    The functions $\psi(x)$, $A(x)$, $U(x)$ and  $U(\psi)$ are plotted in
    \fig 2. One can notice that, in the present example, the potential is
    everywhere negative and is unbounded below; though, its finiteness in
    the outer region of the BH indicates that the latter property is not
    necessary for obtaining BH solutions but is a consequence of our simple
    choice of $R(x)$.

\apsection{B} {A particlelike solution}

    Creation of such an example is a more difficult task than the previous
    one since one now has to take into account two boundary conditions, at
    the centre ($\rho=\rho_c$, $r(\rho_c)=0$) and at the infinity.
    Let us, using the same notations (\ref{dimls}), make the following
    choice:
\beq
      R^2(x) = \frac{1}{u^2} \frac{\tanh (u+c)}{\tanh c},     \label{R-sol}
      \cm  	u\equiv \frac{1}{x},
\eeq
    where $c$ is a positive constant, and put $x_0=0$ in \eq (\ref{A-x}). As
    before, one can explicitly find the quantities $A$ and $\varphi$ in
    terms of $x$ (or $u$) from \eqs (\ref{phi-x}) and (\ref{A-x}):
\bearr
    \sqrt{2}\varphi = \sqrt{3}\arctan \frac{Z}{\sqrt{3}}
        - \Half \log\frac{Z+1}{Z-1}, 			\label{psi-sol}
\nnn \cm
		Z \equiv  \sqrt{2 \cosh [2(u+c)]-1},
\\ \lal                                                   \label{A-sol}
       A = A_0\tanh (u+c)\biggl[1 - \frac{2}{uA_0}
	   + \frac{2}{u} \frac{\sinh u}{\sinh c \cosh (u+c)}
\nnn\inch
	   + \frac{2}{u^2} \log \frac{\sinh (u+c)}{\sinh c}\biggr],
\ear
    where $A_0 = \tanh c$ is the value of $A$ at the centre, $u=\infty$.
    One can verify that the regular centre conditions (\ref{c1}) hold at
    $u=\infty$.

    The expression for $U(x)=a^2 V$, obtained from \eq(\ref{00d}), can also
    be found explicitly, but is too long to be presented here.

    The field $\varphi$ varies between two finite values at large and small
    $x$, and the potential $V$ in this range is smooth and finite too.
    The corresponding plots are presented in \fig 3.

\Picture{82}
{
\unitlength=0.50mm
\special{em:linewidth 0.4pt}
\linethickness{0.4pt}
\begin{picture}(157.00,152.00)
\put(12.00,7.00){\vector(0,1){145.00}}
\put(10.00,42.00){\vector(1,0){70.00}}
\put(10.00,22.00){\makebox(0,0)[rc]{0.8}}
\put(9.00,42.00){\makebox(0,0)[rc]{1}}
\put(10.00,62.00){\makebox(0,0)[rc]{1.2}}
\put(10.00,82.00){\makebox(0,0)[rc]{1.4}}
\put(10.00,102.00){\makebox(0,0)[rc]{1.6}}
\put(10.00,122.00){\makebox(0,0)[rc]{1.8}}
\put(10.00,142.00){\makebox(0,0)[rc]{2}}
\put(32.00,43.00){\makebox(0,0)[cb]{2}}
\put(52.00,43.00){\makebox(0,0)[cb]{4}}
\put(72.00,43.00){\makebox(0,0)[cb]{6}}
\bezier{124}(19.00,113.00)(15.00,140.00)(12.00,138.00)
\bezier{388}(19.00,113.00)(29.00,58.00)(69.00,47.00)
\bezier{48}(12.00,19.00)(15.00,18.00)(20.00,25.00)
\bezier{240}(20.00,25.00)(33.00,40.00)(73.00,39.00)
\put(35.00,78.00){\makebox(0,0)[lb]{$\varphi(x)$}}
\put(35.00,29.00){\makebox(0,0)[lt]{$A(x)$}}
\put(96.00,84.00){\vector(0,1){67.00}}
\put(91.00,146.00){\vector(1,0){66.00}}
\put(116.00,148.00){\makebox(0,0)[cb]{1}}
\put(136.00,148.00){\makebox(0,0)[cb]{2}}
\put(94.00,131.00){\makebox(0,0)[rc]{-0.2}}
\put(94.00,116.00){\makebox(0,0)[rc]{-0.4}}
\put(94.00,101.00){\makebox(0,0)[rc]{-0.6}}
\put(96.00,5.00){\vector(0,1){67.00}}
\put(94.00,52.00){\makebox(0,0)[rc]{-0.2}}
\put(94.00,37.00){\makebox(0,0)[rc]{-0.4}}
\put(94.00,22.00){\makebox(0,0)[rc]{-0.6}}
\put(91.00,67.00){\vector(1,0){66.00}}
\put(116.00,69.00){\makebox(0,0)[cb]{1.4}}
\put(136.00,69.00){\makebox(0,0)[cb]{1.8}}
\put(146.00,69.00){\makebox(0,0)[cb]{2}}
\bezier{100}(96.00,96.00)(100.00,94.00)(106.00,114.00)
\bezier{188}(106.00,114.00)(115.00,147.00)(128.00,146.00)
\bezier{308}(110.00,67.00)(137.00,69.00)(143.00,19.00)
\put(116.00,37.00){\makebox(0,0)[cc]{$U(\varphi)$}}
\put(124.00,110.00){\makebox(0,0)[cc]{$U(x)$}}
\end{picture}
}
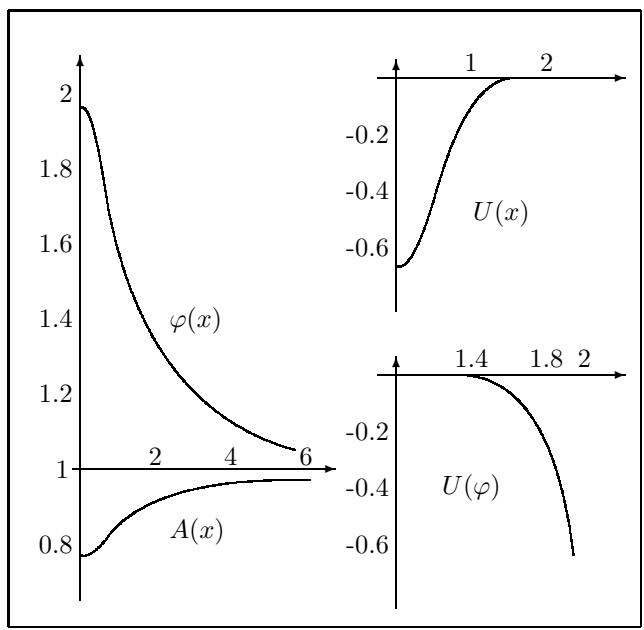
{The functions $\varphi(x)$, $A(x)$, $U(x)$ and  $U(\varphi)$
 for the example of a particlelike solution with $c=1$.}

    The \Sch\ mass, calculated as before, is connected with $c$ or $A_0$ as
    follows:
\beq
	M = \frac{a}{6} \frac{1-A_0^2}{A_0} = \frac{a}{3\sinh(2c)}.
\eeq
    where $a$ is the arbitrary length scale introduced in (\ref{dimls}).

    Thus we have constructed a family of solitonic solutions with an
    arbitrary positive mass.

\apsection{C} {Conformal scalar field in GR: \bh{}s and wormholes}

    Conformal scalar field in GR can be viewed as a special case of STT,
    such that, in \eq (\ref{act-J}), $D=4$ and
\beq
    f(\phi) = 1 - \phi^2/6, \qquad  h(\phi)=1, \qquad U(\phi) =0.
\eeq
    After the conformal mapping
\bear                                                       \label{g-con}
    g\mn \eql F(\varphi)\og\mn, \qquad
     				F(\varphi)=\cosh^2(\varphi/\sqrt{6}),
\\
     \phi \eql \sqrt{6} \tanh (\varphi/\sqrt{6}),           \label{phi-con}
\ear
    we obtain the action (\ref{act-d}) with $D=4$ and $V\equiv 0$. The
    latter describes a minimally coupled massless scalar field in GR, and
    the corresponding \ssph\ solution is well-known: it is the Fisher
    solution \cite{fisher}.  It is convenient to write it using the harmonic
    radial coordinate $u$ specified by the condition \cite{br73} $|g_{uu}| =
    g_{tt}g_{\theta\theta}^2$ (in the previous notation, $u$ behaves as
    $1/r$ at large $r$):
\bear
    ds^2_{\rm E} \eql \e^{-2mu}dt^2
     		- \frac{k^2\e^{2mu}}{\sinh^2(ku)}
        \biggl[\frac{k^2 du^2}{\sinh^2(ku)} + d\Omega^2\biggr],
\nn
     \varphi \eql \sqrt{6}C (u+u_0),                       \label{fish}
\ear
    where the subscript ``E'' stands for the Einstein frame,
    $m$ (the mass), $C$ (the scalar charge), $k>0$ and $u_0$ are
    integration constants, and $k$ is expressed in terms of $m$ and $C$:
\beq
	k^2 = m^2 + 3C^2.                                 \label{k-fish}
\eeq
    The previously used coordinate $\rho$, corresponding to the metric
    (\ref{ds-d}), $D=4$, is connected with $u$ as follows:
\beq
	\rho = k \cot (ku),                                \label{rho-u}
\eeq
    and the metric in terms of $\rho$ has the form
\bearr                                                   \label{ds-fish}
    ds^2_{\rm E}= \biggl(\frac{\rho-k}{\rho+k}\biggr)^{m/k} dt^2
\nnn \cm
    - \biggl(\frac{\rho+k}{\rho-k}\biggr)^{m/k}
          	  \biggl[d\rho^2 + (\rho^2 - k^2)d\Omega^2\biggr].
\ear

    This solution is asymptotically flat at $u\to 0$ ($\rho\to\infty$), has
    no horizon when $C\ne 0$ (as should be the case according to the no-hair
    theorem) and is singular at the centre ($u\to \infty$, $\rho\to k$,
    $\varphi\to\infty$). It turns into the \Sch\ solution when $C=0$.

    The ``Jordan-frame'' solution is described by the metric
    $ds^2 = F(\phi) ds_{\rm E}^2$ and the $\phi$ field according to
    (\ref{g-con}). It is the conformal scalar field solution \cite{bbm70,
    bek74}, whose properties are more diverse and can be enumerated as
    follows (putting, for definiteness, $m>0$ and $C>0$):

\medskip\noi
{\bf 1.} $C < m$. The metric behaves qualitatively as in the Fisher
    solution:  it is flat at $u\to 0$, and both $g_{tt}$ and
    $r^2=|g_{\theta\theta}|$ vanish at $u\to \infty$ --- a singular
    attracting centre. A difference is that here the scalar field is finite:
    $\phi\to \sqrt{6}$.

\medskip\noi
{\bf 2.} $C > m$. Instead of a singular centre, at $u\to\infty$ one has a
    repulsive singularity of infinite radius: $g_{tt}\to \infty$
    and $r^2 \to \infty$. Again $\varphi\to \sqrt{6}$ as $u\to \infty$.

\medskip\noi
{\bf 3.} $C = m$. In this case the metric and the scalar field are regular
    at $u = \infty$, and a continuation across this regular sphere may be
    achieved using a new coordinate, e.g.,
\beq
	y=\tanh (mu).                                         \label{u-y}
\eeq
    The solution acquires the form
\bear
     ds^2 \eql (1+yy_0)^2\biggl[\frac{dt^2}{(1+y)^2}
\nnn \inch
		-\frac{m^2(1+y)^2}{y^4(1-y_0)^2}(dy^2+y^2 d\Omega^2)\biggr],
\nn
     \phi \eql \sqrt{6} \frac{y+y_0}{1 + yy_0},          \label{con-y}
\ear
     where $y_0 = \tanh(m u_0)$. The range $u\in \R_+$, which describes the
     whole manifold $\oM$ in the Fisher solution, corresponds in $\M$ to the
     range $0 < y < 1$, describing only a region of $\M$, the manifold of
     the solution (\ref{con-y}). The properties of the latter depend on the
     sign of $y_0$ \cite{br73}. In all cases, $y=0$ corresponds to a flat
     asymptotic, where $\phi \to \sqrt{6}y_0$, $|y_0| < 1$.

\medskip\noi
{\bf 3a:} $y_0 < 0$. The solution is defined in the range $0 < y <1/|y_0|$.
     At $y=1/|y_0$, there is a naked attracting central singularity:
     $g_{tt}\to 0$, $r^2\to 0$, $\phi\to\infty$.

\medskip\noi
{\bf 3b:} $y_0 > 0$. The solution is defined in the range $y\in \R_+$.
     At $y\to\infty$, we find another flat spatial infinity, where
     $\phi\to \sqrt{6}/y_0$, $r^2\to\infty$ and $g_{tt}$ tends to a finite
     limit.  This is a wormhole solution, found for the first time by one
     of the authors \cite{br73} and recently discussed by Barcelo and
     Visser \cite{viss}.

\medskip\noi
{\bf 3c:} $y_0=0$, $\phi=\sqrt{6}y$, $y\in \R_+$. In this case it is helpful
     to pass to the conventional coordinate $r$, substituting $y=m/(r-m)$.
     The solution
\bear
     ds^2 \eql (1-m/r)^2{dt^2} - \frac{dr^2}{(1-m/r)^2} -r^2d\Omega^2,
\nn
     \phi \eql \sqrt{6}m/(r-m)                              \label{con-bh}
\ear
     is the well-known BH with a conformal scalar field \cite{bbm70,bek74},
     which seems to violate the no-hair theorem. The infinite value of
     $\phi$ at the horizon $r=m$ does not make the metric singular since, as
     is easily verified, the energy-momentum tensor remains finite there.

     The whole case 3 belongs to variant III in the classification of \sect
     5.2, and the horizon in case 3c is situated in the region
     $\M_2 = \M \setminus \M_1$, where the action of the no-hair theorem
     cannot be extended.

     In case 3b, the second spatial infinity and even the wormhole throat
     ($y= 1/\sqrt{y_0}$) are situated in $\M_2$, illustrating the inferences
     of \sect 5.

     However, in case 2, where the mapping is type I by the same
     classification ($\M \map\ \oM$), there appears a minimum of $r(u)$ in
     the metric $g\mn$ (\ref{g-con}), and $r$ even blows up at large $u$.
     This is connected with blowing up of the conformal factor $F$.
     Recall that in \sect 5.2 the absence of another spatial infinity was
     only guaranteed under the finiteness condition for the conformal
     factor in the whole range of the radial coordinate, including its
     boundary values: we see that this condition is indeed essential.

     The simple example of the conformal field thus illustrates the possible
     nontrivial consequences of conformal mappings.

\subsection*{Acknowledgements}

This work was supported in part by the Russian Foundation for Basic
Research. KB acknowledges partial financial support from NATO Science
Fellowship grant and kind hospitality at the Dept. of
Mathematics at the University of the Aegean, Karlovassi, Samos, Greece. We
are grateful to Spiros Cotsakis, Oleg Zaslavskii, Mikhail Katanaev, Olaf
Lechtenfeld and Thomas Strobl for helpful discussions.

\end{document}